\documentclass[twoside,reqno]{heron}
\usepackage{epsfig,cite,colordvi}
\usepackage{graphicx}
\usepackage{url}
\usepackage{amssymb,amsmath,amscd,epsf}
\usepackage{times}
\usepackage{makeidx}
\newcommand{\el}{\left}
\newcommand{\er}{\right}

\newcommand{\vrr}{\varrho}

\newcommand{\veps}{\varepsilon}
\newcommand{\vphi}{\varphi}

\newcommand{\dis}{\displaystyle}

\begin{document}

\title{Estimation of the breakup cross sections in  $^6$He+$^{12}$C reaction within high-energy approximation and
microscopic optical potential}

\runningheads{Estimation of the breakup cross sections in  $^6$He+$^{12}$C...}
{{E.V.~Zemlyanaya}, {V.K.~Lukyanov}, {K.V.~Lukyanov}}

\begin{start}

\author{E.V.~Zemlyanaya}{},
\coauthor{V.K.~Lukyanov}{},
\coauthor{K.V.~Lukyanov}{}

\address{Joint Institute for Nuclear Research, Dubna 141980,
Russia}  {}


\begin{Abstract}
The breakup cross sections in the  reaction $^6$He+$^{12}$C are calculated
at about 40 MeV/nucleon using the high-energy approximation (HEA) and with the help of
microscopic optical potentials (OP) of interaction with the target nucleus $^{12}$C of
the projectile nucleus fragments $^4$He and 2n. Considering the di-neutron $h$=2n as a single
particle the relative motion $h\alpha$ wave function is estimated so that to explain both
the separation energy of $h$ in $^6$He and the rms radius of the latter.  The stripping and
absorbtion total cross sections are calculated and their sum is compared with the total reaction
cross section obtained within a double-folding microscopic OP for the $^6$He+$^{12}$C scattering.
It is concluded that the breakup cross sections contribute in about 50$\%$ of the total reaction
cross section.
\end{Abstract}

\end{start}

\section*{Introduction}
In recent calculations \cite{LuKa2010}, the data on elastic scattering of $^6$He on $^{12}$C at comparably large energies 38.3 and 41.6 MeV/nucleon \cite{Lapoux2002},\cite{Khalili96} were studied using the microscopic optical potentials (OP) \cite{Lukyanov2004}, whose depths
of real and imaginary parts as well as the strength of the surface term were corrected by the three fitted re-normalization coefficients  N$_R$, N$_I$ and N$^{sf}_I$. It was shown that because of the limited set of experimental data the ill-posed problem reveals itself, and therefore
not one but the number of sets of adjusted  N's (and the respective OP's)
were obtained, each characterized by fairly small $\chi^2$ value. In this connection, the study of physics of the process is desirable, namely the search of details of mechanism of the $^6$He+$^{12}$C interaction in different channels. At this stage we intend to study  constituents of a total reaction cross section $\sigma_R$ , the breakup $\sigma_b$ and absorption $\sigma_a$ cross section, and compare them with $\sigma_R$ obtained with the help of the aforementioned OP's in elastic channel.
%
\section{The model of $^6$He}
We consider the simplest breakup $h\alpha$-model of $^6$He, where it is suggested consisting of two  clusters $^4$He and $h$, the correlated pair of neutrons $h$=2n (the similar model was also treated in \cite{Khalili96}). The interaction between clusters is taken to be a WS potential with the adjusted geometrical parameters $R=1.45\, fm$, $a=0.3\,fm$ and the depth $V_0=28.3\, Mev$ that reproduces the separation energy $\veps=0.975 MeV$ of $h$  and yields the $r_{rms}$ radius $2.62\, fm$ of $^6$He. The obtained s-wave function $\vphi_b({\bf s})$ of relative motion of clusters  defines the density distribution
\begin{equation}\label{eq1}
\vrr_b(s) \,=\,|\vphi_b({\bf s})|^2\,=\,(1/4\pi)|\vphi_{l=0}(s)|^2
\end{equation}
and  will be used for the further calculations of the ground state matrix elements of breakup processes. Figure 1 exhibits that $\vrr_b(r)$,
normalized to 1, coincides fairly well with $\vrr_{L}(s)$, the nucleon density distribution of $^6$He obtained within the known large-scale shell-model \cite{Karataglidis2000} (LSSM-model) which also gives $r_{rms}=2.586\, fm$. Thus, we may apply the 2-cluster $h\alpha$-model for the further calculations of elastic and breakup cross sections.
%
%
\section{Folding potentials}
In the framework of the $h\alpha$-model of $^6$He one can estimate the $^6$He+$^{12}$C OP as folding of two OP's of interaction of clusters $\alpha$ and $h$ with $^{12}$C:
\begin{equation}\label{eq2}
\begin{array}{l}
U^{(b)}_{HeC}(r)= V^{DF\,(b)}+iW^{(b)} \, = \\
\qquad =\, \int d^3s\,\vrr_b(s) \Bigl\{U_{\alpha}\bigl({\bf r}-(2/3){\bf s}\bigr)+U_{h}\bigl({\bf r}+(1/3){\bf s}\bigr)\Bigr\} \, = \\
\qquad =\, 2\pi\int\limits_0^\infty \vrr_b(s) s^2ds \int\limits_{-1}^1dx\,\biggl\{ U_\alpha\el(\sqrt{r^2+(1/9)s^2-r(2/3)sx}\,\er) \, +\\
\qquad\qquad +\, U_h\el(\sqrt{r^2+(4/9)s^2+r(4/3)sx}\,\er)\biggr\}.
\end{array}
\end{equation}
Here the $h$-$^{12}$C potential is taken as the twice neutron-$^{12}$C OP $U_h=2U_n$. In turn, potentials  $U_\alpha$ and $U_n$ are calculated within the microscopic hybrid model of OP \cite{Lukyanov2004}. In the latter, the double-folding (DF) real part $V^{DF}$ is constructed as is done in  \cite{Khoa2000},\cite{Kostya2007}, while the imaginary part is derived using the optical limit of a Glauber theory. So, the real and imaginary
parts of OP are as follows:
$$
V^{DF}(r)=V^D(r)~+~V^{EX}(r)= \int d^3 s_p d^3 s_t \Bigl \{ \vrr_p({\bf s}_p)\,\vrr_t({\bf s}_t)\, v_{NN}^D(s)\, +
$$
\begin{equation}\label{eq3}
+\,\int d^3 s_p \, d^3 s_t \, \vrr_p({\bf s}_p, \,{\bf s}_p+ {\bf
s}) \, \vrr_t({\bf s}_t, {\bf s}_t-{\bf s}) v_{NN}^{EX}(s)\,
\exp\el[{i{\bf K}(r)\cdot \frac s M}\er]\Bigr\},
\end{equation}
\begin{equation}\label{eq4}
W^H(r)\,=\,-\frac{1}{2\pi^2}\frac{E}{k}{\bar\sigma}_{N} \,
\int_0^{\infty} j_0(q r){\vrr}_p(q){\vrr}_t(q){f}_N(q) q^2dq.
\end{equation}
Here $p$ and $t$ are related to the projectile and target nucleus, ${\bf s}={\bf r}+{\bf s}_t-{\bf s}_p$, $M=A_pA_t/(A_p+A_t)$, $K(r)$ is the local nucleus-nucleus momentum, and ${\bar\sigma}_N$,the total NN cross section, averaged over the isospins of colliding nuclei. The current calculations
apply the $v_{NN}$ effective Paris nucleon-nucleon CDM3Y6 potentials (for details see in \cite{Khoa2000},\cite{Kostya2007}). As to the density distributions  we use the two-parameter symmetrized fermi-densities $\vrr_p$ and $\vrr_t$ for nuclei $^4$He and $^{12}$C from \cite{LukyanovZemSlow2004}. Thus, $U_{\alpha}$ and $U_h=2U_n$ OP's have the form
\begin{equation}\label{eq5}
U_i(r)\,=\,V_i^{DF}(r)\,+\,iW_i(r),\qquad i=\alpha,~h,
\end{equation}
where $W(r)$ is either  $W^{H}(r)$ or $V^{DF}(r)$. Substituting OP's of fragments (\ref{eq5}) in eq.(\ref{eq2}), the respective real $V^{DF\,(b)}$
and imaginary $W^{(b)}$ parts of OP for $^6$He+$^{12}$C scattering are taken as results of folding with the $h\alpha$-model wave function. These parts are applied to construct the whole $^6$He+$^{12}$C OP as follows
\begin{equation}\label{eq6}
U^{opt\,(b)}_{HeC}\, = \, N_R V^{DF\,(b)}(r) + i N_I W^{(b)}(r),
\end{equation}
where the coefficients $N_R$ and $N_I$ are adjusted to get agreement with the respective experimental data on elastic scattering differential cross sections.
%
%
\begin{figure}
\begin{center}
\includegraphics[ height = 2.7in, width = .8\linewidth]
{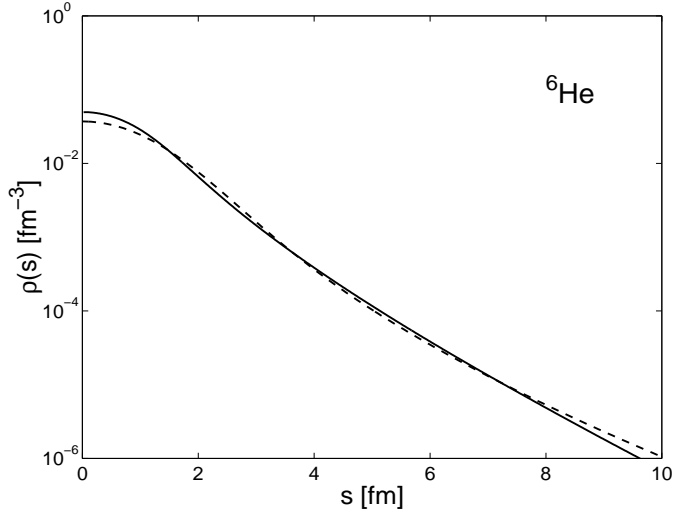}
\end{center}
\caption{\sl Comparison of the $h\alpha$-model density distribution $\vrr_b(s)$ (solid) with the
LSSM density \cite{Karataglidis2000} (dashed). }
\end{figure}
%
%
\begin{figure}[t]
\begin{center}
\includegraphics[ height = 2.7in, width = .8\linewidth]
{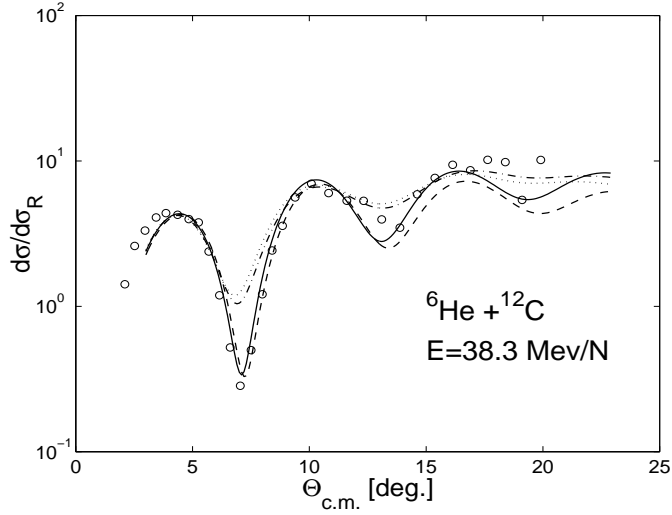}
\end{center}
\caption{\sl The $^6$He+$^{12}$C differential elastic cross sections at 38.3 MeV/N
calculated using $\vrr_b$ density of the $h\alpha$-model for folding OP (eqs.(2),(6)):
solid curve - for W$^{(b)}$=W$^{H\,(b)}$, dashed - for W$^{(b)}$=V$^{DF\,(b)}$.
Dash-dotted and dotted curves are the entire double-folding calculations from \cite{LuKa2010} with
the LSSM nucleon density of ~ $^6$He and with W=V$^{DF}$ and W=W$^{H}$, respectively (eqs.(3),(4)).
The re-normalization N's coefficients are in Table 1. Experimental data from \cite{Lapoux2002}.}
\end{figure}
%
%
\begin{figure}[t]
\begin{center}
\includegraphics[ width = .99\linewidth]
{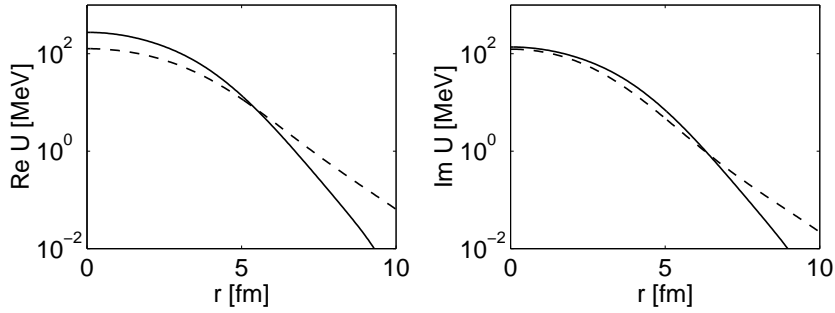}
\end{center}
\caption{\sl The $h\alpha$-model potential for $^6$He+$^{12}$C  elastic scattering at E=38.3 MeV/nucleon (solid) in comparison with the entire DF microscopic OP's applied in \cite{LuKa2010} (dashed). Left panel: real part; right panel: imaginary part. }
\end{figure}
%
\section{Elastic scattering}
Doing so, we apply $U^{opt\,(b)}_{HeC}(r)$ (\ref{eq6}) to consider elastic scattering of $^6$He
from $^{12}$C at E=38.3 MeV/nucleon. In this case, there were applied two kinds of OP, with imaginary parts W=W$^{H\,(b)}$ and
W=V$^{DF\,
(b)}$, and the corresponding differential cross sections were numerically calculated using the code DWUCK4 \cite{DWUCK}.
Besides, we compare these results with cross sections given in \cite{LuKa2010} where the entire double-folding OP (\ref{eq3}) was
utilized accounting for the LSSM density for $^6$He \cite{Karataglidis2000}, and for the $^{12}$C density from \cite{LukyanovZemSlow2004}.
Comparisons were made with the experimental data from \cite{Lapoux2002}. The fitted re-normalization coefficients N's are shown in
Table 1. One can see from Fig.2 that angular distributions for different kinds of ImOP in the $h\alpha$-model (solid and dashed curves)
as well as in the entire DF-model (dash-dotted and dotted curves) are closely displayed, and  the corresponding total
reaction cross sections are almost equal in value as seen from Table 1. Also,  Fig.3 shows the resulting $^6$He+$^{12}$C optical
potentials, that correspond to the case of selection of the HEA $Im$OP (\ref{eq4}) used in the $h\alpha$-model and in the entire
DF-model. One sees that the $Im$OP for both models are rather similar. Nevertheless, we note that the sharper slope
in the periphery of the $h\alpha$-model OP's leads to the pronounced angular distributions as compared to those calculated within the
smooth DF-potential based on the $^6$He LSSM density. As a whole the $h\alpha$-model of $^6$He seems to be reliable for the further
evaluations of total breakup cross sections, that is the subject of our study in the paper.
\begin{table}[t]
\caption{The adjusted \{N\} coefficients of OP and the DWUCK calculations within $h\alpha$- and DF-models
for elastic cross sections in Fig.2.}
\begin{center}
\begin{tabular}{|l|l|l|l|}
\hline
potential & $N_R$ & $N_I$ & $\sigma^{tot}_R$,~mb \\
\hline
$h\alpha$-model, $N_RV^{DF(b)}+iN_IW^{H(b)}$ & & & \\
solid, eq.(6)                               &   2.0  & 1.7 & 1018  \\
\hline
$h\alpha$-model, $N_RV^{DF(b)}+iN_IV^{DF(b)}$& & & \\
 dashed, eq.(6) &  2.1   &  1.0  & 1042 \\
\hline
entire DF-model,  $N_RV^{DF}+iN_IW^{H}$& & & \\
 dotted, ref.[1] &  1.268 &  0.511 & 1029\\
\hline
entire DF-model,  $N_RV^{DF}+iN_IV^{DF}$ & & & \\
dash-dotted, ref.[1] & 1.123   &   0.472  & 1034  \\
\hline
\end{tabular}
\end{center}
\end{table}
%
\section{Testing the HEA(eikonal) method}
For calculations of breakup cross sections, 
the analytic eikonal (HEA) method is utilized. As to our further applications of HEA approach at energies of about 40 MeV/nucleon we should preliminary verify that
this method is well working. For this purpose we calculate the notably characteristic of a process, the differential cross section of the $^6$He+$^{12}$C elastic scattering at 38.3 MeV/nucleon, within the numerical code DWUCK4 and also using the HEA method. In both cases we
apply the same microscopic double-folding OP $U^{opt}=(1.123 +i0.472)V^{DF}(r)$ from \cite{LuKa2010}. For this OP the exact result for the
angular distribution was already shown in Fig.2 by the dashed-dotted curve. As to the analogical eikonal calculations we first
exhibit the explicit expression for the HEA amplitude of scattering (for details see ref.\cite{LukZem2001})
\begin{equation}\label{eq7}
f(q)=f_{pc}(q)+ik\int_0^\infty db~bJ_0(qb)~{\dis e}^{-{\dis i\Phi_{pc}}} \el(1-
{\dis e}^{\dis i\Phi_N+i\delta\Phi_{uc}}\er),
\end{equation}
where $q=2k\sin(\vartheta/2)$ is the transfer momentum, $f_{pc}(q)$, the known amplitude of scattering in the field of  the Coulomb potential $U_{pc}=Z_1Z_2e^2/r$. Then, $\delta \Phi_{uc}=\Phi_{uc}-\Phi_{pc}$ is the difference of eikonal phases for the potential of a uniformly charged sphere and the $U_{pc}$ potential, while the nuclear eikonal phase is
\begin{equation}\label{eq8}
\Phi_N=-{\frac k
E}\int\limits_0^\infty\,U^{opt}_{HeC}\el(\sqrt{b^2+z^2}\er)dz.
\end{equation}
Note that when performing integration in (\ref{eq7}) the
trajectory distortion is taken into account by exchanging the
impact parameter $b$ by the distance of closest approach in the
Coulomb field $U_{pc}$ at $b$=0, i.e. $b\rightarrow b_c={\bar
a}+\sqrt{{\bar a}^2+b^2}$ with ${\bar a}=Z_pZ_te^2/2E_{c.m.}$.

In Fig.4 is shown the comparison of two curves for $d\sigma/d\sigma_R$ where $d\sigma_R$ is the Rutherford cross section for scattering in the $U_{pc}$ potential. The solid curve corresponds to the HEA method, and the dashed one is the exact DWUCK calculations.  One can see that both curves coincide fairly well, especially  at small angles, in the region that yields the main contribution to the total cross sections. Thus we conclude that the HEA method may be applied for our further estimations of the total breakup cross sections.
\begin{figure}[t]
\begin{center}
\includegraphics[ height = 2.7in, width = .8\linewidth]
{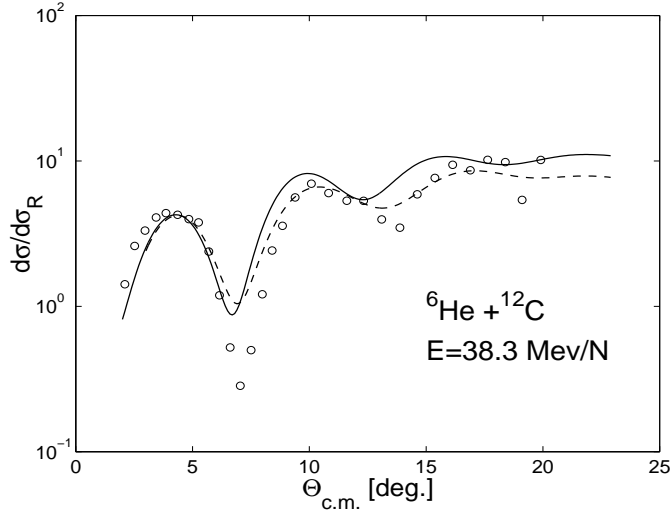}
\end{center}
\caption{\sl Differential cross sections of the $^6$He+$^{12}$C
elastic scattering at E=38.3 MeV/nucleon calculated for the same $U_{opt}=(1.123+i0.472)\cdot V^{DF}$
from \cite{LuKa2010} by using the eikonal method (solid curve) and the DWUCK4 code (dashed curve).}
\end{figure}
%
\section{The HEA model for breakup reactions}
The earlier HEA theory for the breakup processes were developed in refs.\cite{Glauber1955},\cite{Akhiezer1955} for investigations of stripping and dissociation of deuterons in nuclear collisions. In
recent papers (see, e.g.,  \cite{Bertsch1996},\cite{Bertulani2004} and refs therein) this method was generalized
to study breakup reactions of lightest nuclei. For a brief review of this method we begin with the conditions $E\gg|U|$, $\vartheta\ll
(1/kR)^{1/2}$ when the OP wave function of a high-energy particle can be considered in the eikonal form:
\begin{equation}\label{eq9}
\Psi({\bf r})\,=\,{\dis e}^{\dis i{\bf k}{\bf r}-{\frac i{\hbar v}
}\int_{-\infty}^z dz \, U^{opt}\el(\sqrt{b^2+z^2}\er)},
\end{equation}
After scattering at z$\rightarrow+\infty$ this function becomes
\begin{equation}\label{eq10}
\Psi({\bf r})\,=\,S(b)\cdot{\dis e}^{\dis i{\bf k}{\bf r}}, \qquad
S(b)\,=\,{\dis e}^{\dis -{\frac i {\hbar v}}\int_{-\infty}^\infty
dz \, U^{opt}\el(\sqrt{b^2+z^2}\er)},
\end{equation}
where $S(b)$ is an analog of the partial $S_l$-matrix, and formulae defined by $S_l$ may be transformed to respective
expressions with $S(b)$ using relations $l+1/2\rightarrow kb$ and $(1/k)\sum_l\rightarrow\int db$. So, after the
collision the probability that the particle with an impact parameter $b$ remains in the elastic channel is
\begin{equation}\label{eq11}
|S_i(b)|^2\,=\,{\dis e}^{\dis -{\frac 2{\hbar v}}\int_0^\infty dz \,
W_i\el(\sqrt{b^2+z^2}\er)},\qquad i=\alpha, h,
\end{equation}
and the probability for the particle to be removed from the elastic channel is $(1- |S|^2)$. (Here we denote $W=|{\rm Im} U|$.)
Thus, the common probability of both $h$
and $\alpha$ particles to leave the elastic channel is $(1- |S_h|^2)(1- |S_\alpha|^2)$. Then, one should average this latter by $\vrr_b(s)$
that characterizes the probability of $h$ and $\alpha$ to be at relative distance $s$. As a result, for the $h\alpha$-model of $^6$He
{\it the total  absorbtion cross section} is obtained as follows
\begin{equation}\label{eq12}
\sigma_{abs}^{tot}\,=\,2\pi~ \int_0^\infty b_hdb_h\el(1\,-\,|S_h(b_h)|^2\er)\,\el(1\,-\,I(b_h)\er),
\end{equation}
where
\begin{equation}\label{eq13}
I(b_h)=\int d^3s \vrr_b(s)|S_\alpha(b_\alpha)|^2, \quad b_\alpha=\sqrt{s^2\sin^2\vartheta + b_h^2- 2sb_h\sin\vartheta\cos\phi}
\end{equation}
Here the relation is used  of impact parameters ${\bf b}_\alpha={\bf b}_h-{\bf b}$ with $b=s\sin\vartheta$ being the projection of
the $h-\alpha$ vector ${\bf s}$ on the plane normal to the 0z-axis along the straight line trajectory of an incident nucleus.\\
In the case of the stripping reaction with removing $h$-particle from $^6$He to the target nucleus, one should use the probability
of $h$ to leave the elastic channel $(1-|S_h(b_h)|^2)$, and for $\alpha$ to continue its elastic scattering with probability
$|S_\alpha(b_\alpha)|^2$. Then the  probability of the whole process is $|S_\alpha(b_\alpha)|^2\cdot(1-|S_h(b_h)|^2)$, and to
get the total stripping cross section one must average over $\vrr_b(s)$ as is done in (\ref{eq12}),(\ref{eq13}). In a similar
manner the transfer of the $\alpha$ particle can be constructed,
and the net contribution of both removal reactions yields  {\it
the total breakup cross section}
\begin{equation}\label{eq14}
\sigma_{bu}^{tot}\,=\,2\pi \,\int_0^\infty b_hdb_h \Bigl \{|S_h(b_h)|^2+\el[1\,-\,2|S_h(b_h)|^2\er]\cdot I(b_h)\Bigr \}.
\end{equation}
The sum of the absorption (\ref{eq12}) and breakup (\ref{eq14}) cross sections results in {\it the total reaction cross section}
\begin{equation}\label{eq15}
\sigma_R^{tot}\,=\,2\pi~ \int_0^\infty b_hdb_h\Bigl ( 1\,-\,|S_h(b_h)|^2 \cdot I(b_h)\Bigr )
\end{equation}
\begin{table}
\caption{The HEA estimations within the $h\alpha$-model of total cross sections of $^6$He+$^{12}$C at E=38.3 MeV/nucleon.}
\begin{center}
\begin{tabular}{|l|l|l|l|}
\hline
potential &	$\sigma_{abs}^{tot}$, mb &	$\sigma_{bu}^{tot}$, mb & $\sigma_{R}^{tot}$, mb  \\
\hline
$Im$OP=N$_I$W$^{H(b)}$,  eq.(4),  N$_I$=1.7 &	392	& 412	& 804 \\
\hline
$Im$OP=N$_I$V$^{DF(b)}$, eq.(3), N$_I$=1.0	& 447	& 389	& 830 \\
\hline
\end{tabular}
\end{center}
\end{table}
%
\section{Summary and conclusions}
Estimations of the total cross sections  were made with a help of the preliminary calculated imaginary
parts of optical potentials $U_h$ and $U_\alpha$ for scattering of $h$- and $\alpha$-particles on $^{12}$C. Firstly, we treated them as the $N_IW_{h,\alpha}^H$ potential done by eq.(\ref{eq4}) of HEA, and also, in the other attempt, they were taken  in the double-folding form $N_IV^{DF}$ eq.(\ref{eq3}) usually used for the real potentials. The re-normalization coefficients $\{N_I\}$ are the same as they were fitted for the folded potentials (\ref{eq2}) of the $h\alpha$ model (Table 1, rows 2,3). Thereafter the respective probabilities  of scattering  $|S_{h,\alpha}|^2$ (\ref{eq11}) were obtained and applied in calculations of the respective cross sections (\ref{eq12}),(\ref{eq14}),(\ref{eq15}) shown in Table 2.
One can see that in this case the total reaction cross sections $\sigma^{tot}_R = 804,\, 830\, mb$ turn out to be about 20\% lower than those
$\sigma^{tot}_R = 1018,\, 1042 \, mb$ obtained within the code DWUCK4 for the $U_{HeC}^{(b)}$ optical potential (\ref{eq2}), the result of folding
the $U_{h}$ and $U_{\alpha}$  potentials with the $h\alpha$-density function $\vrr_b(s)$.
This 20\% difference seems not too large, but to get the more substantial conclusion
one should make comparisons of results forfolded OPs calculated not within the code DWUCK4 but using the eikonal  expression for the total
reaction cross section \cite{Glauber1955}
\begin{equation}\label{eq16}
\sigma_R^{tot}=2\pi~ \int_0^\infty bdb\el\{1- \exp\el [-{\frac 2 {\hbar v}}\int_{-\infty}^\infty dz \,W\el(\sqrt{b^2+z^2}\er )\er ]\er\}.
\end{equation}
One should underline that here it is involved only the imaginary parts $W$ of the $U_{HeC}^{(b)}$ optical potential (\ref{eq2}), in our case they are
$W=1.7 W^H$ and  $W=1.0 V^{DF}$. So, using (\ref{eq16}) we got the respective reaction cross sections $\sigma_R^{tot}$= 952 and 965 $mb$.
Thus, the difference of these results from the preceding HEA results $\sigma^{tot}_R$ = 804 and 830 $mb$ is only about 10\%. The small rest
discrepancy can arise due to the additional role of the real part of OP in the DWUCK calculations, while the HEA expression (\ref{eq16})
depends only on the imaginary part of OP.
The other effect is  ought to the difference in formulae (\ref{eq15}) and (\ref{eq16}). Indeed, in the first one the density $\vrr_b$ folds
in eq.(\ref{eq13}) probability function $|S_{h,\alpha}|^2$ having the bare potential $W$ in the exponent. Otherwise, the cross section (\ref{eq16}) contains the already folded potential in its exponent. By the way these effects occur to be not too significant, and one can conclude that the main mechanism of the absorbtion in elastic channel of the $^6$He+$^{12}$C scattering is ought to existence of the power dissociation channels of the $^6$He in two clusters $h=2n$ and $\alpha$.

\noindent
{\bf Acknowledgments.}~~~The authors would like to thank Prof. A.~Antonov and Prof. W.~Scheid for helpful remarks and suggestions. The work was supported by the
Program for collaboration of JINR and Bulgarian scientific centers. EVZ and KVL thank RFBR (grant No.09-01-00770) for partial financial support.

\end{document}